\documentstyle[aps,prd,floats,psfig]{revtex}
\preprint{DAMTP-2002-xyz}
\date{September 2002}
\tighten
\begin{document}
\title{Cosmological Evolution of Brane World Moduli}
\author{Ph. Brax${}^1$, C. van de Bruck${}^2$, A.--C. Davis${}^{2}$ 
and C.S. Rhodes$^2$}
\address{$^1$Service de Physique Th\'eorique, CEA-Saclay\\ 
F-91191, Gif/Yvette cedex, France \\
and\\
$^2$Department of Applied Mathematics and Theoretical Physics, 
Center for Mathematical Sciences,\\
University of Cambridge, Wilberforce Road, Cambridge CB3 0WA, U.K.} 
\noindent
\maketitle
\begin{abstract}
We study cosmological consequences of non--constant brane world 
moduli in five dimensional brane world models with bulk scalars 
and two boundary branes. We focus on the case where the  brane tension
is an exponential function of the bulk scalar field, $U_b \propto \exp{(\alpha \phi)}$. 
In the limit $\alpha \rightarrow 0$, the model reduces to the two--brane 
model of Randall--Sundrum, whereas larger values of $\alpha$ allow
for a less warped bulk geometry. Using the moduli space approximation we 
derive the four--dimensional low--energy effective action from a 
supergravity--inspired five--dimensional theory. 
For arbitrary values of $\alpha$, the resulting theory has the form 
of a bi--scalar--tensor theory. We show that, in order to be 
consistent with local gravitational observations, 
$\alpha$ has to be small (less than $10^{-2}$) and the 
separation of the branes must be large.  
We study the cosmological evolution of the interbrane 
distance and the bulk scalar field for different matter contents 
on each branes. Our findings indicate that attractor solutions exist
which drive the moduli fields towards values consistent with observations.
The efficiency of the attractor mechanism crucially depends on the 
matter content on each branes. In the five--dimensional description, 
the attractors correspond to the motion of the negative tension brane 
towards a bulk singularity, which signals the eventual breakdown of
the four--dimensional 
description and the necessity of a better understanding of the bulk
singularity. 
\end{abstract}

\vspace{0.25cm}

\noindent DAMTP--2002--108; T02/108

\vspace{0.15cm}

\section{Introduction}
One of the most exciting ideas which has originated from theories of 
particle physics beyond the Standard Model is that our universe could
be a three--dimensional object (called a brane) embedded in a
higher--dimensional space (the bulk). Special attention has been paid
to models in which the bulk is five--dimensional. For these models,
cosmological consequences have been worked out in particular for the
cases of heterotic M--theory
--- a five--dimensional theory with two branes, which are the
boundaries of the bulk spacetime (see
e.g. [\ref{Lukas1}-\ref{Lukas4}]) --- and the Randall--Sundrum model
[\ref{RS1},\ref{RS2}], in both the version with one brane and two
branes (see e.g. [\ref{Bine1}-\ref{Cline2}]).  In the case of the
latter model, the bulk spacetime is highly warped (in fact, the bulk
is an Anti--de Sitter spacetime), whereas in heterotic M--theory the
bulk is ``slightly'' warped. Another essential ingredient in heterotic
M--theory is the existence of a bulk scalar field, whereas in the
original Randall--Sundrum model there is no such bulk scalar. It was
later introduced in order to stabilize the inter--brane distance
[\ref{Wise1},\ref{Wise2}].

It was shown in several papers that the resulting four--dimensional
theories at low energies in both models have much in common with
(multi)scalar--tensor theories, where the interbrane distance and the
bulk scalar field degree of freedom play the r\^{o}le of scalar
degrees of freedom in the gravitational sector of the effective
four--dimensional theory [\ref{Garriga1}-\ref{Kanno2}].  From the
cosmological point of view it is interesting to investigate the
consequences of these moduli. Although it is usually assumed that
these moduli fields are stabilized by some unknown mechanism in the
early universe [\ref{Choi}-\ref{Steinhardt}], it is imaginable that
they are not stabilized by some potential. Instead, it might be that
there is a cosmological attractor mechanism at work, similar to the
one found in Brans--Dicke type theories, which drives the theory
towards four--dimensional General Relativity during the cosmological evolution by
generating very small matter couplings of the moduli
[\ref{Damour1}-\ref{Wagoner}].  If this is the case, there could be
cosmological consequences of the time--evolution of the moduli, such
as varying constants, time--varying masses, etc., leading to
observational consequences, such as in the cosmic microwave background
radiation, large scale structures, time--varying constants,
Equivalence Principle violations, and so on. Indeed, there is another 
motivation for our work: namely, the claim of a time--varying 
fine--structure constant made in [\ref{webb1}] and 
[\ref{webb2}]. If this claim is confirmed by future observations, 
this would suggest that at least some moduli fields are not 
stabilized but slowly varying. 

In this paper we investigate the cosmological evolution of moduli
fields arising in brane world models and address the question of
whether there is a cosmological attractor for the brane world moduli.
We  also analyse the conditions under  which conditions such an attractor would be efficient enough
for the theory to agree with local experiments in the laboratory
and/or strong field limits [\ref{Will}]. We study  these issues
in a broad class of models encapsulating features from both the
Randall--Sundrum models and heterotic M--theory, i.e.
a warped background with a bulk scalar field.

The paper is organized as follows: in the next section we discuss the
five--dimensional theory and derive in detail the effective
low--energy action from the moduli space approximation, giving 
also the resulting field equations in the Einstein frame.  In Section 3
we clarify the conditions under which the theory would predict
time--varying coupling constants, such as the fine--structure
constant, and investigate the constraints imposed by strong field
limits.  Cosmological considerations and solutions to the field
equations are presented in Section 4, and we conclude in Section 5.

\section{Moduli space approximation}
In this section we derive the four--dimensional low--energy effective action
using the moduli space approximation.  For the gravitational sector,
our derivation follows closely that in [\ref{Garriga2}]. We begin with
the description of the static configuration, which was derived from
supergravity [\ref{Brax2}].

\subsection{The static configuration}
The bulk action consists of two terms which describe 
gravity and the bulk scalar field dynamics:
\begin{equation}
S_{\rm bulk} = \frac{1}{2\kappa_5^2} \int d^5 x \sqrt{-g_5}
\left( {\cal R} - \frac{3}{4}\left( (\partial \psi)^2 + U \right)\right).
\end{equation}
Further, our setup contains two branes. One of these branes has a
positive tension, the other brane has a negative tension.  They are
described by the action
\begin{eqnarray}
S_{\rm brane 1} &=& -\frac{3}{2\kappa_5^2}\int d^5x \sqrt{-g_5} U_B
\delta(z_1), \label{b1} \\
S_{\rm brane 2} &=& +\frac{3}{2\kappa_5^2}\int d^5x \sqrt{-g_5} U_B \delta(z_2) \label{b2}.
\end{eqnarray}
In these expressions, $z_1$ and $z_2$ are the (arbitrary) positions of
the two branes, $U_B$ is the superpotential; $U$, the bulk potential
energy of the scalar field, is given by
\begin{equation}
U = \left(\frac{\partial U_B}{\partial \psi}\right)^2 - U_B^2.
\end{equation}
We will also include the Gibbons--Hawking boundary term for each
brane, which have the form
\begin{equation}
S_{\rm GH} = \frac{1}{\kappa_5^2}\int d^4 x \sqrt{-g_4} K,
\end{equation}
where $K$ is the extrinsic curvature of the individual branes.
We impose a $Z_2$--symmetry at the position of each brane. 

The solution of the system above can be derived from 
BPS--like equations of the form
\begin{equation}
\frac{a'}{a}=-\frac{U_B}{4},\ \psi'=\frac{\partial U_B}{\partial \psi},
\end{equation}
where $'=d/dz$ for a metric of the form
\begin{equation}\label{background}
ds^2 = dz^2 + a^2(z)\eta_{\mu\nu}dx^\mu dx^\nu.
\end{equation}
We will particularly focus on the case where the superpotential is an 
exponential function:
\begin{equation}\label{potential}
U_B=4k e^{\alpha \psi}.
\end{equation}
The values  $\alpha =1/\sqrt 3,-1/\sqrt {12}$ were obtained in a theory
with supergravity in singular spaces [\ref{Brax2}]. The solutions read
\begin{equation}\label{scale}
a(z)=(1-4k\alpha^2z)^{\frac{1}{4\alpha^2}},
\end{equation}
while the scalar field solution is
\begin{equation}\label{psi}
\psi = -\frac{1}{\alpha}\ln\left(1-4k\alpha^2z\right).
\end{equation}
In the $\alpha\to 0$ we retrieve the AdS profile
\begin{equation}
a(z)=e^{-kz}.
\end{equation}
Notice that in that case the scalar field decouples altogether. Also, 
notice that there is a singular point in the bulk at 
$z_* = 1/4k\alpha^2$, for which the scale factor vanishes. This will be 
important for the discussion on the cosmological evolution in Section 4. 

In the following we will discuss the moduli space approximation.  We
will put matter on the branes as well as supersymmetry breaking
corrections to the brane potentials. Two of the moduli of the
system are the brane positions.  That is, in the solution above the
brane positions are arbitrary.  In the moduli space approximation,
these moduli are assumed to be space--time dependent.  We denote the
position of brane 1 with $z_1 = \phi(x^\nu)$ and the position of
brane 2 with $z_2 = \sigma(x^\mu)$. We consider the case where the
evolution of the brane is slow. This means that in constructing the
effective four--dimensional theory we neglect terms like $(\partial
\phi)^3$.

In addition to the brane positions, we need to include the graviton 
zero mode, which can be done by replacing $\eta_{\mu\nu}$ with a 
space--time dependent tensor $g_{\mu\nu}(x^\mu)$. Thus, we have 
two scalar degrees of freedom, namely the positions of the two 
branes which we will denote with $\phi(x^\mu)$ and $\sigma(x^\mu)$, 
and the graviton zero mode $g_{\mu\nu}$ (see [\ref{us}] and 
[\ref{cynolter}]). As we will see below, in 
the coordinate system in which the branes move, the kinetic terms 
of the moduli come from the boundary terms alone. 

It would also be possible to consider another coordinate system, in
which the branes are fixed. Then, the moduli are a space--time
dependent part which can be added to the bulk scalar field solution
(\ref{psi}) and the 55--component of the metric tensor becomes a
four--dimensional effective field. In addition, we have the graviton
zero--mode. Thus, the number of moduli fields is independent of the
coordinate system.

Note that the moduli space approximation is only a good approximation
if the time--variation of the moduli fields is small. This should be
the case for late--time cosmology well after nucleosynthesis, which we 
are interested in (in [\ref{Turok1}] and [\ref{Turok2}] the moduli
space approximation was also used in the context of brane worlds).

\subsection{Moduli space approximation: the gravitational sector}
Let us first consider the bulk action. Replacing $\eta_{\mu\nu}$ 
with $g_{\mu\nu}(x^{\mu})$ in (\ref{background}) we have for the 
Ricci scalar $R = R^{(4)}/a^2 + \tilde R$, where $\tilde R$ is the 
Ricci--scalar of the background (\ref{background}). We explicitly use
the background solution (\ref{scale}) and (\ref{psi}), so that 
$\tilde R$ will not contribute to the low--energy effective action. 
Also, in this coordinate system, where the branes move, there is no 
contribution from the part of the bulk scalar field. Collecting 
everything we therefore have
\begin{equation}
S_{\rm bulk} = \frac{1}{2\kappa_5} \int dz d^4 x a^4
\sqrt{-g_4}\frac{1}{a^2}{\cal R}^{(4)} = \int d^4 x 
\sqrt{-g_4} f(\phi,\sigma) {\cal R}^{(4)},
\end{equation}
with 
\begin{equation}
f(\phi,\sigma) = \frac{1}{\kappa_5^2} \int^{\sigma}_{\phi} dz a^2 (z).
\end{equation}
We remind the reader that $a(z)$ is given by (\ref{scale}). 

We now turn to the boundary terms. It is clear that the integrals 
(\ref{b1}) and (\ref{b2}) do not contribute to the effective action 
for the same reason that $\tilde R$ does not contribute. Let us 
therefore turn our attention to the Gibbons--Hawking boundary terms.

First, it is not difficult to construct the normal vectors to the
brane:
\begin{equation}
n^\mu = \frac{1}{\sqrt{1+ (\partial \phi)^2/a^2}}\left( \partial^\mu
\phi/a^2,1\right).
\end{equation}

Then the induced metric on each brane is given by
\begin{eqnarray}
g_{\mu\nu}^{\rm ind,1} &=& a^2(\phi) g_{\mu\nu}^4 - \partial_{\mu}\phi
\partial_{\nu} \phi, \\
g_{\mu\nu}^{\rm ind,2} &=& a^2(\sigma) g_{\mu\nu}^4 - \partial_{\mu}\sigma
\partial_{\nu} \sigma.
\end{eqnarray}
Thus, 
\begin{eqnarray}
\sqrt{-g^{\rm ind,1}} &=& a^4(\phi) \sqrt{-g_4}\left[ 1 -
\frac{1}{2a^2(\phi)}(\partial \phi)^2\right], \\
\sqrt{-g^{\rm ind,2}} &=& a^4(\sigma) \sqrt{-g_4}\left[ 1 -
\frac{1}{2a^2(\sigma)}(\partial \sigma)^2\right].
\end{eqnarray}
So the Gibbons--Hawking boundary terms take the form 
\begin{eqnarray}
\frac{1}{\kappa_5^2} \int d^4 x a^4 \sqrt{-g_4}\left[ 1 -
\frac{1}{2a^2(\phi)} (\partial \phi)^2 \right] K, \\
\frac{1}{\kappa_5^2} \int d^4 x a^4 \sqrt{-g_4}\left[ 1 -
\frac{1}{2a^2(\sigma)} (\partial \sigma)^2 \right] K.
\end{eqnarray}
The trace of the extrinsic curvature tensor can be calculated from 
\begin{equation}
K = \frac{1}{\sqrt{-g_5}}\partial_\mu \left[ \sqrt{-g_5}n^\mu \right].
\end{equation}
Neglecting higher order terms this gives
\begin{eqnarray}
K = 4\frac{a'}{a}\left[1 - \frac{(\partial \phi)^2}{4a^2}\right].
\end{eqnarray}
The terms for the second brane can be obtained analogously.
Using the BPS conditions and keeping only the kinetic terms, we get
for the Gibbons--Hawking boundary terms
\begin{eqnarray}
&+&\frac{3}{4\kappa_5^2}\int d^4x \sqrt{-g_4} a^2(\phi) U_B(\phi)
(\partial \phi)^2, \\
&-&\frac{3}{4\kappa_5^2}\int d^4x \sqrt{-g_4} a^2(\sigma) U_B(\sigma)
(\partial \sigma)^2.
\end{eqnarray}
Collecting all terms we find
\begin{eqnarray}
S_{\rm MSA} = \int d^4 x \sqrt{-g_4}\left[ f(\phi,\sigma) {\cal R}^{(4)} 
+ \frac{3}{4}a^2(\phi)\frac{U_B(\phi)}{\kappa_5^2}(\partial \phi)^2 
- \frac{3}{4} a^2(\sigma)\frac{U_B}{\kappa_5^2}(\sigma)(\partial \sigma)^2 \right].
\end{eqnarray}
Note that the kinetic term of the field $\phi$ has the wrong
sign. This is an artifact of the frame we use here. As we shall show
below, it is possible to go to the Einstein frame with a simple
conformal transformation, in which the sign in front of the kinetic
term is correct for both fields.

The effective action is valid for any model based on supergravity--inspired
five dimensional models with two branes. In particular it is worth noticing that there is no potential for the moduli. The moduli describe flat directions
reflecting the no-force condition on boundary branes for BPS-like systems. 

The effective action for the two moduli $\phi$ and $\sigma$ has a nice
interpretation in terms of supergravity. This is expected as we
started from a two-brane system satisfying BPS conditions. The
resulting effective action has a supergravity structure.
Indeed one can write the Einstein-Hilbert term and the kinetic terms
of the moduli as
\begin{equation}
 \int d^4 x \sqrt{-g_4}\left[ f(\phi,\sigma) {\cal R}^{(4)} 
+ \frac{1}{6}\partial_{\Phi}\partial_{\bar\Phi}f\partial_\mu
\Phi\partial^\mu \bar \Phi + \frac{1}{6}\partial_{\Sigma }\partial_{\bar\Sigma}f\partial_\mu
\Sigma \partial^\mu \bar \Sigma\right ],
\end{equation}
where
\begin{equation}
\phi=\frac{1}{2}(\Phi+\bar\Phi),\
\sigma=\frac{1}{2}(\Sigma+\bar\Sigma).
\end{equation}
In terms of supergravity, the fields $\Phi$ and $\Sigma$ are the
scalar parts of two chiral multiplets. One can read off the Kahler
potential for these two moduli fields as
\begin{equation}
K=-3\ln f.
\end{equation}
In the Randall--Sundrum case, we retrieve that
\begin{equation}
K=-3\ln ( e^{-k(\Phi+\bar\Phi)}-e^{-k(\Sigma+\bar\Sigma)}).
\end{equation}
Notice that in that case one can rewrite the Kahler potential as
\begin{equation}
K=-3k(\Phi+\bar\Phi)-3\ln(1-e^{-k(T+\bar T)}),
\end{equation}
where
\begin{equation}
T=\Sigma - \Phi
\end{equation}
is the radion superfield.  Moreover we immediately see that, in the
Randall--Sundrum case, the field $\Phi$ can be eliminated by a Kahler
transformation; this shows that one of the two moduli decouples,
leaving only the radion as the relevant physical field. We will
retrieve this result later when we analyse the equations of motion.

Coming back to the action above, we redefine the fields in the following way:
\begin{eqnarray}
\tilde \phi^2 &=& \left(1 - 4k\alpha^2 \phi\right)^{2\beta}, \label{posia1}\\
\tilde \sigma^2 &=& \left(1-4k\alpha^2 \sigma\right)^{2\beta} \label{posia2},
\end{eqnarray}
with 
\begin{equation}
\beta = \frac{2\alpha^2 + 1}{4\alpha^2};
\end{equation}
then, the gravitational sector can be written as
\begin{eqnarray}
S_{\rm MSA} &=&\frac{1}{2k\kappa_5^2(2\alpha^2 + 1)}\int d^4 x \sqrt{-g_4}\left[ \left(\tilde\phi^2 -
\tilde\sigma^2 \right) {\cal R}^{(4)} +
\frac{6}{2\alpha^2 + 1}\left( (\partial \tilde\phi)^2
-(\partial \tilde\sigma)^2\right)\right].
\end{eqnarray}

This is a action of the form of a multi--scalar tensor theory, in
which one scalar field has the wrong sign in front of the kinetic
term. Furthermore, in this frame there is a peculiar point where the
factor in front of ${\cal R}$ can vanish, namely when
$\tilde\phi=\tilde\sigma$, which corresponds to colliding branes. We
will call the frame in which the action has the form of the equation
above the {\it bulk frame}.

It is useful to go to the Einstein frame. In order to avoid mixed
terms like $(\partial_\mu \tilde\phi)(\partial^\mu \tilde\sigma)$, we
shall define two new fields\footnote{Do not confuse the Ricci scalar
${\cal R}$ with the new field $R$.}:
\begin{eqnarray}
\tilde \phi &=& Q \cosh R, \label{posib1} \\
\tilde \sigma &=& Q \sinh R \label{posib2}.
\end{eqnarray}
To go to the Einstein frame we perform a conformal transformation:
\begin{equation}
\tilde g_{\mu\nu} = Q^2 g_{\mu\nu}.
\end{equation}
Then [\ref{Wald}]
\begin{equation}
\sqrt{-g} Q^2 R = \sqrt{-\tilde g} \left( \tilde R -
\frac{6}{Q^2}(\tilde\partial Q)^2 \right) .
\end{equation}
Collecting everything we get the action in the Einstein frame
(where we now drop the tilde):
\begin{eqnarray}
S_{\rm EF} &=& \frac{1}{2k\kappa^2_5(2\alpha^2 + 1)} 
\int d^4x \sqrt{-g}\left[ {\cal R} -  \frac{12\alpha^2}{1+2\alpha^2}
\frac{(\partial Q)^2}{Q^2} - \frac{6}{2\alpha^2 + 1}(\partial R)^2\right].
\end{eqnarray}
Clearly, in this frame both fields have the correct sign in front of
the kinetic terms. Note that for $\alpha \rightarrow 0$ 
(i.e.\ the Randall--Sundrum case) the $Q$--field decouples. 
In this case, the field $R$ plays the r\^{o}le of the radion, i.e. it 
measures the distance between the branes. Furthermore, we can identify 
the gravitational constant:
\begin{equation}
16\pi G = 2k\kappa_5^2 (1+2\alpha^2).
\end{equation}

\subsection{Moduli space approximation: the matter sector}
In the following we introduce matter as well as supersymmetry breaking 
potentials $V(Q,R)$ and $W(Q,R)$ on each branes. We begin with the potentials:
to first order in the moduli space approximation we get
\begin{equation}
\int d^4 x \sqrt{-g_4} \left[ a^4(\phi)V(\phi) \right]
\end{equation}
with $a^4(\phi) = \tilde \phi^{4/(1+2\alpha^2)}$. The expression for 
a  potential $W$ on the second brane is similar with 
$a(\phi)$ is replaced by $a(\sigma)$. In the Einstein frame we have
(dropping the tilde from the metric):
\begin{equation}
\int d^4 x\sqrt{-g} Q^{-8\alpha^2/(1+2\alpha^2)}(\cosh
R)^{4/(1+2\alpha^2)} V(Q,R) \equiv \int d^4 x\sqrt{-g} V_{\rm eff}(Q,R),
\end{equation}
where we have defined 
\begin{equation}
V_{\rm eff}(Q,R) = Q^{-8\alpha^2/(1+2\alpha^2)}(\cosh
R)^{4/(1+2\alpha^2)} V(Q,R).
\end{equation}
The expression for $W(Q,R)$ in the Einstein frame is
\begin{equation}
\int d^4 x\sqrt{-g} Q^{-8\alpha^2/(1+2\alpha^2)}(\sinh
R)^{4/(1+2\alpha^2)} W(Q,R) \equiv \int d^4 x\sqrt{-g} W_{\rm eff}(Q,R),
\end{equation}
where
\begin{equation}
W_{\rm eff}(Q,R) = Q^{-8\alpha^2/(1+2\alpha^2)}(\sinh
R)^{4/(1+2\alpha^2)} W(Q,R).
\end{equation}
For matter, the action has the form
\begin{equation}
S_m^{(1)} = S_m^{(1)}(\Psi_1,g^{B(1)}_{\mu\nu}) \hspace{0.5cm} {\rm and}
\hspace{0.5cm} S_m^{(2)} = S_m^{(2)}(\Psi_2,g^{B(2)}_{\mu\nu}),
\end{equation}
where $g^{B}$ denotes the {\it induced} metric on each branes and
$\Psi_i$ the matter fields on each branes. Note that we do not 
couple the matter fields $\Psi_i$ to the bulk scalar field, and thus 
not to the fields $Q$ and $R$. In going to the Einstein frame we get 
\begin{equation}
S_m^{(1)} = S_m^{(1)}(\Psi_1,A^2(Q,R)g_{\mu\nu}) \hspace{0.5cm} {\rm and}
\hspace{0.5cm} S_m^{(2)} = S_m^{(2)}(\Psi_2,B^2(Q,R)g_{\mu\nu}),
\end{equation}
In this expression we have used the fact that, in going to the
Einstein frame, the induced metrics on each branes transform with a
different conformal factor, which we denote with $A$ and $B$. We have neglected the derivative terms in the moduli fields when considering the coupling to matter on the brane. They lead to higher order operators which can be easily incorporated.  
The energy--momentum tensor in the Einstein frame is defined as
\begin{equation}
T_{\mu\nu}^{(1)} = 2 \frac{1}{\sqrt{-g}} \frac{\delta
S_m^{(1)}(\Psi,A^2(Q,R)g)}{\delta g^{\mu\nu}}
\end{equation}
with an analogous definition for the energy--momentum tensor for
matter on the second brane.

In the Einstein frame, the total action is therefore, 
\begin{eqnarray}
S_{\rm EF} &=& \frac{1}{16\pi G} 
\int d^4x \sqrt{-g}\left[ {\cal R} -  \frac{12\alpha^2}{1+2\alpha^2}
\frac{(\partial Q)^2}{Q^2} - \frac{6}{2\alpha^2 + 1}(\partial
R)^2\right] \nonumber \\
&-& \int d^4 x\sqrt{-g} (V_{\rm eff}(Q,R)+W_{\rm eff}(Q,R)) 
+ S_m^{(1)}(\Psi_1,A^2(Q,R)g_{\mu\nu}) + S_m^{(2)}(\Psi_2,B^2(Q,R)g_{\mu\nu}).
\end{eqnarray}

The theory derived has a form similar to the one discussed in 
[\ref{Gibbons},\ref{Gundlach}], although we have here two scalar
degrees of freedom in general. However, these scalar fields couple
differently to the two matter types on each individual branes. 

\subsection{The field equations in the Einstein frame}
From the action derived above we can now derive the field 
equations for the metric and the two scalar degrees of freedom 
in the Einstein frame. They are
\begin{eqnarray}
G_{\mu\nu} &=& 8\pi G \left( T_{\mu\nu}^{(1)} + T_{\mu\nu}^{(2)} -
g_{\mu\nu}V_{\rm eff} - g_{\mu\nu}W_{\rm eff}\right)
+\frac{12\alpha^2}{1+2\alpha^2}\frac{1}{Q^2} 
\left[ (\partial_\mu Q)(\partial_\nu Q) 
- \frac{1}{2}g_{\mu\nu}(\partial Q)^2 \right] \nonumber\\
&+& \frac{6}{1+2\alpha^2}\left[ (\partial_\mu R)(\partial_\nu R) 
- \frac{1}{2}g_{\mu\nu}(\partial R)^2 \right] \label{Einstein} \\ 
\Box R &=& 8 \pi G \frac{1+2\alpha^2}{6}\left[\frac{\partial V_{\rm eff}}{\partial R} 
+ \frac{\partial W_{\rm eff}}{\partial R} - \alpha_R^{(1)} T^{(1)} -
\alpha_R^{(2)} T^{(2)}\right] \label{Req} \\
\frac{\Box Q}{Q^2} - \frac{(\partial Q)^2}{Q^3} &=& 
 8 \pi G \frac{1+2\alpha^2}{12\alpha^2} \left[\frac{\partial V_{\rm eff}}{\partial Q} 
+ \frac{\partial W_{\rm eff}}{\partial Q} - \alpha_Q^{(1)} T^{(1)} - 
\alpha_Q^{(2)} T^{(2)}\right]\label{Qeq}.
\end{eqnarray}
In these expression $T^{(i)}$ is the trace of the energy--momentum
tensor for each brane's matter; the coupling functions
$\alpha_Q^{(i)}$ and $\alpha_R^{(i)}$ are defined as
\begin{eqnarray}
\alpha_Q^{(1)} &=& \frac{\partial \ln A}{\partial Q}, \hspace{0.5cm} \alpha_Q^{(2)} = \frac{\partial \ln B}{\partial Q};\\
\alpha_R^{(1)} &=& \frac{\partial \ln A}{\partial R}, \hspace{0.5cm} \alpha_R^{(2)} = \frac{\partial \ln B}{\partial R}.
\end{eqnarray}
We will give the expressions for these quantities in the next section 
and discuss solutions to the field equations in a cosmological 
setting in Section 4. Due to the coupling of the moduli to the 
branes, cosmological matter is not conserved but satisfies
\begin{equation}
D_{\mu}T^{\mu\nu}_i = \alpha_Q^i(\partial^\nu Q)
T_i + \alpha_R^i(\partial^\nu R) T_i
\end{equation}
for each type of matter $i=1,2$. In deriving this equation we have
assumed that the matter fields obey the equation of motion. 
Notice that matter non--conservation is directly linked to 
the space-time variations of the moduli fields.

\section{Couplings to matter and strong field constraints}
In this section we will be interested in the coupling of matter living
on both branes to the bulk. In particular we will pay particular
attention to the constraints imposed by strong field limits. This, in
turn, leads to stringent constraints on the parameter $\alpha$ and 
the allowed values of the field $R$. 

Let us first notice that matter on the branes couple to the induced
metric. The form of the induced metric implies that we have to deal
with a bi-metric theory. This is a general result for moving branes.
In the following we will only be interested in the non-derivative
couplings between the two moduli and the matter fields. Indeed we are
dealing with a low energy expansion and derivative interaction lead to
higher order operators. If need be these derivative interactions can
easily incorporated in the following.

First of all couplings to the two branes results from a Lagrangian
\begin{equation}
\int \sqrt{-g_B} {\cal L}_m(\phi_m,g_B),
\end{equation}
where $\phi_m$ can either be a scalar, fermion or vector boson
field. We will study each case separately.
Let us start with a scalar field, i.e. the Lagrangian is given by 
\begin{equation}
{\cal L}_m=
\frac{1}{2}g_B^{\mu\nu}\partial_\mu\phi_B\partial_\nu\phi_B -V(\phi_B).
\end{equation}
We first write the boundary action in terms of the Einstein
frame metric $\tilde g= \Omega^{-2} g$ leading to 
\begin{equation}
\int \sqrt{-\tilde g}\left( a^2(\phi)\Omega^2 \tilde
g^{\mu\nu}\partial_\mu\phi_B\partial_\nu\phi_B
-a^4(\phi)\Omega^4V(\phi_B)\right),
\end{equation}
where $\Omega^2=1/f(\phi,\sigma)$.
Now we can redefine the scalar field $\tilde \phi= a(\phi)\Omega
\phi_B$ in such a way that the action becomes
\begin{equation}
\int \sqrt{-\tilde g}\left( \tilde g^{\mu\nu}\partial_\mu\tilde
\phi\partial_\nu\tilde \phi -a^4(\phi)\Omega^4 V\left(\frac{\tilde
\phi}{a(\phi)\Omega}\right)\right),
\end{equation}
up to derivative interactions.
Let us apply this to a simple renormalizable potential
\begin{equation}
V=\frac{1}{2}m^2 \phi_B^2 + \lambda \phi^4.
\end{equation}
In the Einstein frame, the couplings become
\begin{equation}
\tilde m= a(\phi)\Omega m,\ \tilde \lambda=\lambda.
\end{equation}
Clearly, in this frame the mass of the scalar field becomes
time--dependent.

We can apply the same technique to vector bosons with action
\begin{equation}
S_m=-\frac{1}{4g^2}\int\sqrt{-g_B}g_B^{\mu\rho}g_B^{\nu\lambda}F_{\mu\rho}F_{\nu\lambda}
\end{equation}
We find that in the Einstein frame
\begin{equation} \label{coupling}
g_E = g,
\end{equation}
as expected vector bosons are conformally coupled and therefore the
gauge coupling constant are time-independent.
Finally let us deal with Dirac fermions:
\begin{equation}
S_m= \int \sqrt{-g_B}( i\bar \psi \gamma_B^{\mu}D_\mu\psi-m\bar
\psi \psi ),
\end{equation}
where $\gamma_B^\mu$ are the gamma matrices for the induced metric.
Using the results derived in [\ref{Brax4}], we can rewrite the action
as 
\begin{equation}
S_m= \int \sqrt{-\tilde g}\left( \overline {\tilde\psi} \tilde
\gamma^{\mu}D_\mu\tilde\psi -m a(\phi)\Omega \overline{
\tilde\psi}  \tilde\psi \right),
\end{equation}
where we have used the conformality of the coupling of massless Weyl
fermions and defined $\tilde \psi= (a(\phi)\Omega)^{3/2}\psi$.
Again we find that fermion masses become time-dependent
\begin{equation}
\tilde m = a(\phi)\Omega m,
\end{equation}
Notice that the time dependence of the fermion and scalar masses is
the same. This is not true when comparing different particle 
species on the positive tension brane with species on the 
negative tension brane. However, particles on the negative tension
brane are candidates for dark matter (they couple to the standard 
model particles only via gravity). Constraints imposed on dark 
matter by Equivalence Principle violations are less restrictive.

In conclusion we see that, in the Einstein frame, the only 
time-dependent couplings in the matter Lagrangian are the 
masses. Indeed, in a general frame we have the following 
invariant relating Newton's constant to masses
\begin{equation}\label{invariant}
I=G m^2=\tilde G \tilde m^2.
\end{equation}
which is a dimension-less quantity. In particular we find that in the
frame where masses are time-independent the Newton constant becomes
time-dependent. As such $I$ gives the true measure of time variations 
in the brane-world models that we consider.

The fact that the coupling constants are frame--independent 
(see eq. (\ref{coupling}))   
and therefore space--time independent means, that in the theory we
consider we would not expect a time--varying fine--structure constant, 
as reported in [\ref{webb1},\ref{webb2}]. In order to explain 
a time--varying coupling constant, one needs to couple vector 
bosons directly to the bulk scalar field and thus to $Q$ and $R$. We 
do not consider this possibility in this paper.

Let us now consider the constraints imposed by the strong--field limits.
For that it is convenient to write the moduli Lagrangian in the form
of a non-linear sigma model with kinetic terms
\begin{equation}
\gamma_{ij}\partial \phi^i\partial \phi^j,
\end{equation}
where $i=1,2$ labels the moduli $\phi^1=Q$ and $\phi^2=R$.
The sigma model couplings are here
\begin{equation}
\gamma_{QQ}= \frac{12\alpha^2}{1+2\alpha^2}\frac{1}{Q^2},\
\gamma_{RR}=\frac{6}{1+2\alpha^2}.
\end{equation}
Notice the potential danger of the $\alpha\to 0$ limit, the RS model,
where the coupling to $Q$ becomes very small. In an ordinary
Brans-Dicke theory with a single field, this would correspond to a
vanishing Brans-Dicke parameter which is ruled out
experimentally. Here we will see that the coupling to matter is such
that this is not the case. Indeed we can write the action expressing
the coupling to
ordinary matter on our brane 
as
\begin{equation}  
S_m=S_m( \phi_m,A^2  \tilde g),\  A=a(\phi)\Omega ,
\end{equation}
where we have neglected the derivative interaction and 
expressed\footnote{The parameter $B$ which defines the coupling
of matter to the second brane is similarly defined by
$B=a(\sigma)\Omega$.} $g_B=A\tilde g$.
Notice that $A$ enters  in both the coupling of matter to the brane
and the time variation of masses. As such it represents the
actual time-dependence of couplings in our models.  
Let us introduce the parameters
\begin{equation}
\alpha_Q=\partial_Q \ln A,\ \alpha_R=\partial_R \ln A.
\end{equation}
We find that ($\lambda = 4/(1+2\alpha^2)$) 
\begin{equation}
A=Q^{-\frac{\alpha^2\lambda}{2}}(\cosh R)^{\frac{\lambda}{4}},
\end{equation}
leading to
\begin{equation}
\alpha_Q= -\frac{\alpha^2\lambda}{2}\frac{1}{Q}, \alpha_R=\frac{\lambda
\tanh R}{4}.
\end{equation}
Observations constrain the parameter
\begin{equation}
\theta=\gamma^{ij}\alpha_i\alpha_j
\end{equation}
to be less than $10^{-3}$ [\ref{gilles}].
We obtain therefore a bound on
\begin{equation}
\theta= \frac{4}{3}\frac{\alpha^2}{1+2\alpha^2}+ \frac{\tanh^2 R}{6(1+2\alpha^2)}.
\end{equation}
The bound implies that
\begin{equation}
\alpha\lesssim 10^{-2},\ R\lesssim 0.2
\end{equation}
This already rules out the supergravity model based on $\alpha^2=1/12$
[\ref{Brax2}].  The smallness of $\alpha$ indicates a strongly warped
bulk geometry such as an Anti--de Sitter spacetime.  In the case
$\alpha=0$, we can easily interpret the bound on $R$.  Indeed in that
case
\begin{equation}
\tanh R = e^{-k(\sigma -\phi)},
\end{equation}
i.e. this is nothing but the exponential of the radion field measuring
the distance between the branes. We retrieve the well-known fact that
gravity experiments require the branes to be sufficiently far apart.
When $\alpha\ne 0$ but small, one way of obtaining a small value of $R$
is for the hidden brane to become close from the would-be singularity
where $a(\sigma)=0$. In the following we will analyse the cosmological 
behaviour of the present model and in particular the robustness of the 
condition $R \ll 1$ to cosmological evolution. 

\section{Cosmological evolution of the two brane system}
As we have seen from the last section, the field $R$ has to be small
today in order for the theory to be in agreement with
observations.  This required smallness of $R$ could result from one of
the following possible features of our model: firstly, the minimum of
$V_{\rm eff}$ or $W_{\rm eff}$ could be at small values of
$R$, and consequently $R$ would have been driven towards this
minimum in the very early universe;  the form of the potential has to be
derived from an underlying theory, whose form is unknown. A much more
interesting alternative is that $R$ could be driven towards small
values during the late cosmological evolution, i.e. after
nucleosynthesis, via an attractor mechanism. 
In scalar--tensor theories such a mechanism is well known (see
e.g. [\ref{Damour1}-\ref{Wagoner}]) and the question we address in
this section is whether such an attractor mechanism exists for the
brane world models we have discussed so far. Clearly, in our model
there are {\it two} moduli fields, $Q$ and $R$ and even if $R$ is
rapidly driven towards small values, there is the danger that the
dynamics of the $Q$ field has no attractor behaviour. This could
jeopardize the cosmology of the model we consider. Furthermore, even
if such an attractor mechanism exists, it is not {\it a priori} clear
whether it is efficient enough.

To address these important issues, we now discuss cosmological
solutions to equations (\ref{Einstein}), (\ref{Req}) and
(\ref{Qeq}). In this case the field equations are as follows 
($\alpha \neq 0$): The
Friedmann equation reads
\begin{equation}\label{Friedmann}
H^2 = \frac{8 \pi G}{3} \left(\rho_1 + \rho_2 + V_{\rm eff} 
+ W_{\rm eff} \right) + \frac{2\alpha^2}{1 + 2\alpha^2} \dot\phi^2
+ \frac{1}{1+2\alpha^2} \dot R^2.
\end{equation}
where we have defined $Q=\exp \phi$. 
The second Einstein equation is
\begin{eqnarray}\label{secEinstein}
\dot H + H^2 &=& -\frac{4 \pi G}{3} \left( \rho_1 + 3p_1 
+ \rho_2 + 3p_2  - 2 V_{\rm eff} - 2W_{\rm eff}\right) \nonumber\\
&-& \frac{4\alpha^2}{1+2\alpha^2} \dot\phi^2 
- \frac{2}{1+2\alpha^2} \dot R^2
\end{eqnarray}
The field equations for $R$ and $\phi$ read
\begin{eqnarray}
\ddot R + 3 H \dot R &=& - 8 \pi G \frac{1+2\alpha^2}{6}\left[ 
\frac{\partial V_{\rm eff}}{\partial R} + 
\frac{\partial W_{\rm eff}}{\partial R} + 
\alpha_R^{(1)} (\rho_1 - 3p_1) + 
\alpha_R^{(2)} (\rho_2 - 3p_2) \right] \label{Rcos} 
\end{eqnarray} 
\begin{eqnarray}
\ddot \phi + 3 H \dot \phi =
-8 \pi G \frac{1+2\alpha^2}{12 \alpha^2} \left[ 
\frac{\partial V_{\rm eff}}{\partial \phi} + 
\frac{\partial W_{\rm eff}}{\partial \phi} + 
\alpha_\phi^{(1)} (\rho_1 - 3p_1) + 
\alpha_\phi^{(2)} (\rho_2 - 3p_2) \right].\label{Qcos}
\end{eqnarray}
For the model we consider here, we have 
\begin{eqnarray}
\alpha_\phi^{(1)} &=& -\frac{2\alpha^2}{1+2\alpha^2}, \hspace{0.5cm}
\alpha_\phi^{(2)} = -\frac{2\alpha^2}{1+2\alpha^2}, \label{coupling1} \\
\alpha_R^{(1)} &=& \frac{\tanh R}{1+2\alpha^2}, \hspace{0.5cm}
\alpha_R^{(2)} = \frac{(\tanh R)^{-1}}{1+2\alpha^2}. \label{coupling2}
\end{eqnarray}

We will study numerical solutions of the system in more detail below,
but if we consider matter on the positive tension brane only we can draw 
some conclusions concerning the evolution of the fields $\phi$ and $R$: 

We considerer that throughout the radiation-- and matter--dominated
eras the potentials $V_{\rm eff}$ and $W_{\rm eff}$ are
negligible. This is a similar condition as that imposed in
quintessence models. Then, in the radiation--dominated era, the traces
of the energy--momentum tensors vanish, implying that
\begin{equation}
\dot R=a^{-3}, \dot \phi=a^{-3}
\end{equation}
in such a way that $R$ remains small if it is small initially.
Moreover, there is no change to the time dependence of the scale
factor.

In the matter--dominated era, the equations of motion read
\begin{eqnarray}
H^2&=&\frac{8\pi G_N}{3}\rho_1 +
\frac{2\alpha^2}{3(1+2\alpha^2)}\dot\phi^2 +\frac{1}{1+2\alpha^2}\dot
R^2\nonumber \\
\ddot R+ 3H\dot R&=& -\frac{8\pi G_N}{6}\rho_1 R \approx -\frac{H^2}{2} R,
\nonumber \\
\ddot \phi +3H \dot \phi &=& \frac{8\pi
G_N}{6}\rho_1 \approx \frac{H^2}{2} \nonumber
\end{eqnarray}
together with the conservation equation 
\begin{equation}
\dot \rho_1+3H \rho_1 \approx -2\alpha^2 \rho_1 \dot\phi 
\end{equation}
where we assume that both $\alpha$ and $R_0$ are small to comply with
the tests of the Equivalence Principle. 
Assuming that the energy density of matter dominates the expansion
rate, the solutions to these equation is given by 
\begin{equation}
\rho_1=\rho_0\left(\frac{a}{a_0}\right)^{-3-2\alpha^2/3}
,\ a=a_0\left(\frac{t}{t_0}\right)^{2/3-4\alpha^2/27}
\end{equation}
together with
\begin{equation}
\phi=\phi_0+\frac{1}{3}\ln\frac{a}{a_0},\
R = R_1 \left(\frac{t}{t_0}\right)^{-1/3} + 
R_2 \left(\frac{t}{t_0}\right)^{-2/3}  
\end{equation}
as soon as $t\gg t_0$. Note, that this solution is consistent with our
assumptions on the expansion rate as long as $R$ is small at $t=t_0$.
Notice further that $R$ converges towards zero, hence retrieving
general relativity at late times and that the change of the expansion
rate is of order $\alpha^2$ and therefore small. The solution above is
in good agreement with the numerical solution described below. We
therefore conclude that during the matter--dominated era $R$ is driven
to zero, provided $R$ was small at the beginning of the matter era.

\subsection{Numerical Results}
We now describe numerical solutions of the cosmological equations.
There are different interesting cases to study.  Firstly, all matter
types could live on the positive tension brane; a more intriguing
alternative might be that the dark matter lives on the negative
tension brane, whereas radiation and baryons live on the positive
tension brane. We will discuss these cases below.

\subsubsection{No Potentials:}
We begin the analysis with the assumption that the potentials $V$ and
$W$ are identically zero (in other words, that the brane potentials
are unbroken from their supersymmetric values); thus, we follow the
evolution of the fields during radiation and matter domination. Both
matter and radiation live on the positive tension brane.  The
calculations are made with $\alpha = 0.01$.

\begin{figure}[!ht]
\hspace{5cm}
\psfig{file=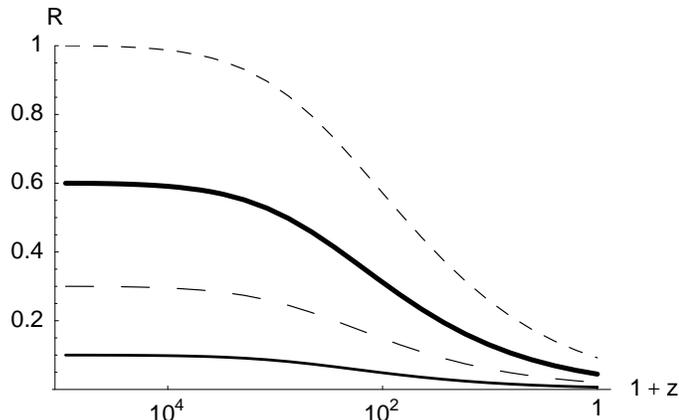,width=9cm}
\caption[h]{The evolution of $R$ with different initial conditions 
for the case of radiation and matter on the 
positive tension brane and no matter on the negative tension brane.
We find that $R$ is driven towards zero.}
\end{figure}

\begin{figure}[!ht]
\hspace{5cm}
\psfig{file=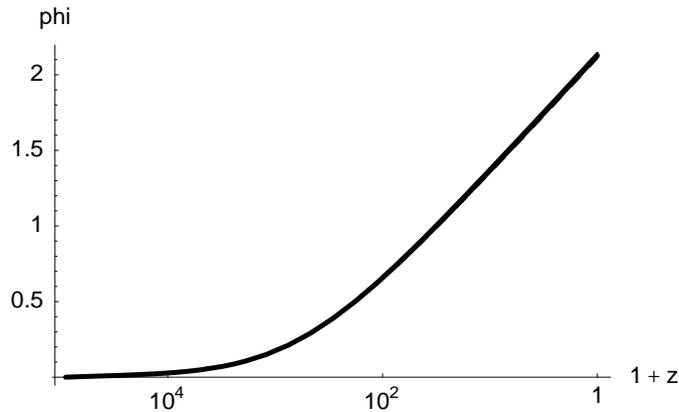,width=9cm}
\vspace{0.75cm}
\caption[h]{The evolution of $\phi$ for different initial 
conditions with the same cosmology as in Figure 1.}
\end{figure}

The evolution of $R$ and $\phi$ are shown in fig. 1 and fig. 2, 
respectively. One can clearly see that during radiation domination 
both fields are frozen in, because the trace of the energy--momentum 
tensor is effectively zero. Soon after matter becomes important, 
both fields are forced to evolve due to the non--vanishing trace of 
the matter energy--momentum tensor. For the initial conditions we have 
chosen the constraint $R<0.2$ can be fulfilled. However, if 
$R$ is initially large, it has not enough time during matter 
domination to evolve to small enough values. Thus, there is 
a constraint that $R$ has to be small enough during radiation 
domination. 

If we put matter on the second brane as well, the evolution of $R$ 
is modified, as can be seen in fig. 3 (see fig. 4 for the evolution
of $\phi$).
For these cases, the matter on the second brane does never 
dominate the expansion of the universe. However, due to the coupling 
function $\alpha_R^{(2)}$, the evolution of $R$ and $\phi$ are 
affected strongly by matter on the second brane. Indeed, as can 
clearly be seen, $R$ is driven faster towards small values even with only 
a small amount of matter on the second brane. The field $\phi$ stays
longer constant after radiation domination if there is matter on the 
second brane.

\begin{figure}[!ht]
\hspace{5cm}
\psfig{file=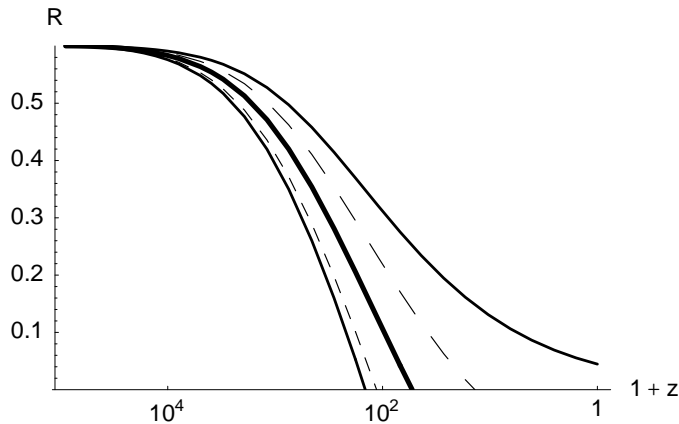,width=9cm}
\vspace{0.75cm}
\caption[h]{The evolution of $R$ with different initial conditions 
for the case of radiation and matter on the 
positive tension brane and pressureless matter on the negative tension brane.
We find that $R$ is driven towards zero for all cases. Note, that
if the ratio $\rho_2/\rho_1$ grows, $R$ evolves faster towards
zero. The ratios of $\rho_2/\rho_1$ are given by 0, 0.11, 0.25, 0.42
and 0.5, for the curves from the right to the left.}
\end{figure}

\begin{figure}[!ht]
\hspace{5cm}\psfig{file=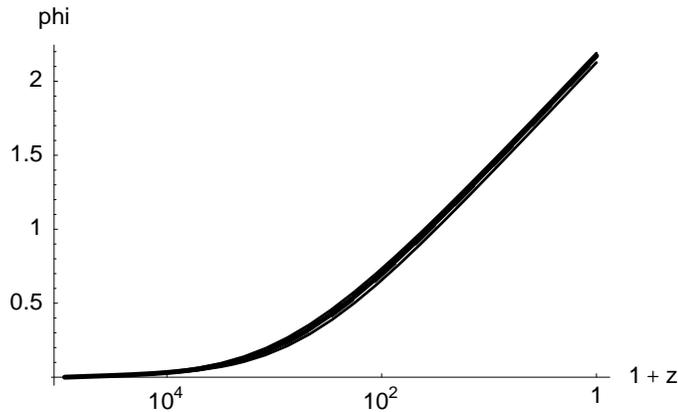,width=9cm}\vspace{0.75cm}
\caption[h]{The evolution of $\phi$ for different initial 
conditions with the same cosmology as in Figure 3.}
\end{figure}

We conclude that there is an cosmological attractor which drives $R$
towards small values and that the efficiency is enhanced if the energy
density of matter on the second brane is not negligible initially. 

\subsubsection{Including Potentials:}
We now include potentials on the branes. We will study the case where
there is only a potential on the positive tension brane. For the 
potential we do assume that it starts to dominate the energy 
density only recently in the cosmic history, because we are 
interested in the evolution of the moduli fields $R$ and $\phi$. 
In principle, both $\phi$ and $R$ are candidates for the dark energy. 
However, we do not address the coincidence problem, nor do we want 
to provide a model for dark energy in this paper (for a discussion 
about dark energy arising in particle physics models, see e.g. 
[\ref{Wetterich}-\ref{Linde}] and references therein). Instead, we use 
the simplest model for the potential, namely we assume that $V$ 
has the form of an exponential potential. To be specific, we assume 
that supersymmetry is broken by tuning $U_{B}$ away from 
the form (\ref{potential}) by setting
\begin{equation}
V = (T-1)4k e^{\alpha \psi}.
\end{equation}
Here, $T\neq 1$ is a supersymmetry breaking parameter ($\psi$ is the
bulk scalar field). Expressed in terms of $\phi$ and $R$ we have
\begin{equation}\label{poti}
V(\phi,R) = 4(T-1)k e^{-12\alpha^2\phi/(1+2\alpha^2)}
\left(\cosh R \right)^{(4-4\alpha^2)/(1+2\alpha^2)}.
\end{equation}
Notice that for $R$ close to zero, this is nothing but an
exponential model with the field $\phi$ playing the role of a
quintessence field.
In the following we will set $4k(T-1)$ such that the universe starts 
to accelerate at a redshift around 1. 

\begin{figure}[!ht]
\hspace{5cm}
\psfig{file=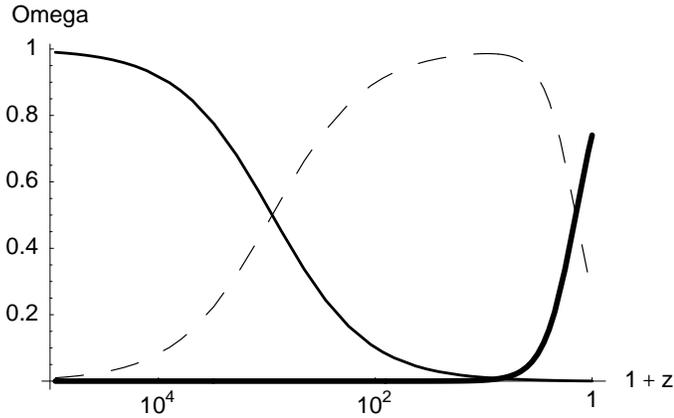,width=9cm}
\vspace{0.75cm}
\caption[h]{Evolution of the density parameter $\Omega_{i}$ as a
function of redshift for radiation, matter and the scalar fields.
When the potential of the moduli dominates, the universe is 
accelerating. Note, that in 
order to explain the values for the energy density of dark energy, 
one has to fine--tune the parameters of the theory. For these plots we
have set $\alpha=0.01$. The dark matter lives on 
the positive tension brane, there is no matter on the negative tension
brane.}
\end{figure}

\begin{figure}[!ht]
\hspace{5cm}
\psfig{file=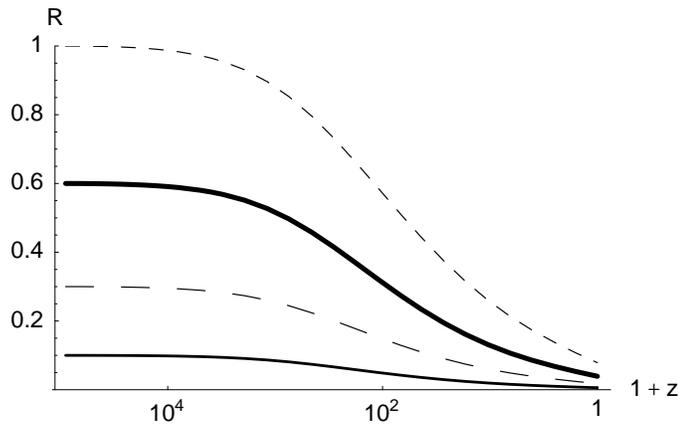,width=9cm}
\vspace{0.75cm}
\caption[h]{Evolution of the field $R$ as a function of redshift 
with different initial conditions. The constraint 
today is $R<0.2$.}
\end{figure}

The evolution of the cosmological parameters is shown in fig. 5. 
After the usual matter dominated era, the universe becomes dominated
by the potential energy of the fields and starts to accelerate. 

The evolution for $R$ is shown in fig. 6. We have chosen the same 
initial conditions for $R$ as in fig. 1 and it can be seen that the 
evolution of $R$ is not much affected by the presence of a potential
which dominates today. In fig. 7 we have plotted the evolution of
$\phi$. Note that as soon as the potential dominates the expansion, 
the evolution of $\phi$ is affected. 

\begin{figure}[!ht]
\hspace{5cm}
\psfig{file=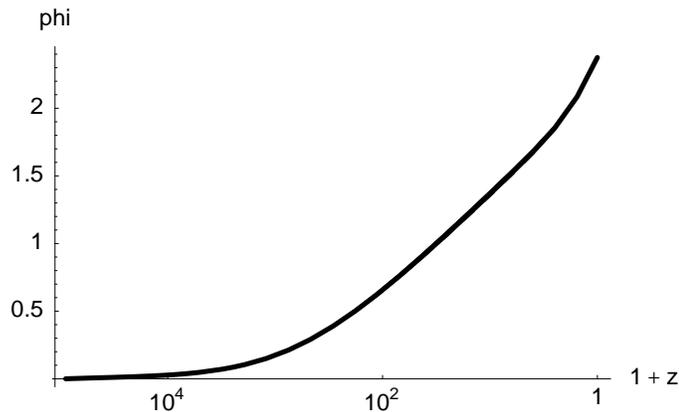,width=9cm}
\vspace{0.75cm}
\caption[h]{Evolution of the field $\phi$ as a function of redshift 
with different initial conditions. Note that when the potential energy
starts to dominate, $\phi$ changes its time--evolution.}
\end{figure}

\subsubsection{Time--variation of Masses or the Gravitational Constant}
As explained in Section 3, the quantity $I$, defined in eq. 
(\ref{invariant}), is an invariant under conformal transformations. 
This quantity specifies the variation of masses in the Einstein frame 
(or the gravitational constant in the frame where the masses on the 
positive tension brane are time--independent). The variation of the 
gravitational constant (and therefore of $I$) is constrained by 
nucleosynthesis (see e.g. [\ref{Uzan}] and [\ref{Chiba2}] for a 
recent discussion). We have plotted the evolution of $I$ for 
different initial conditions, one with initial conditions leading to 
an evolution in agreement with observations (fig. 8) and one extreme case which
is ruled out by nucleosynthesis constraints (fig. 9). Therefore, 
{\it the initial conditions for the fields} $R$ {\it and} $\phi$ 
{\it are not arbitrary, but constrained by nucleosynthesis.} The
details of the evolution of $I$ is strongly dependent on the 
matter contents on the branes. 

\begin{figure}[!ht]
\hspace{5cm}
\psfig{file=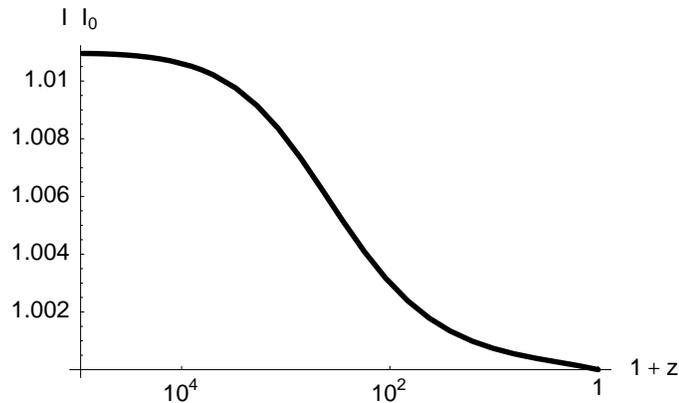,width=9cm}
\vspace{0.75cm}
\caption[h]{Evolution of $I/I_0$ (eq. (\ref{invariant})) for initial 
conditions which are allowed by nucleosynthesis. The value of 
the gravitational constant at nucleosynthesis was of order 
one percent larger than its value today.}
\end{figure}

\begin{figure}[!ht]
\hspace{5cm}
\psfig{file=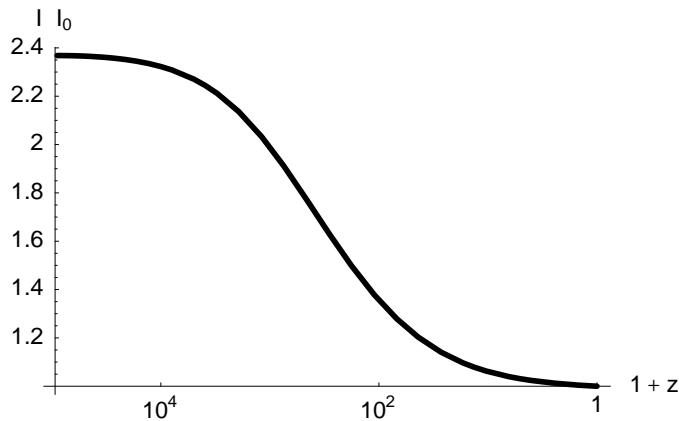,width=9cm}
\vspace{0.75cm}
\caption[h]{Evolution of $I/I_0$ (eq. (\ref{invariant})) for initial 
conditions which are forbidden by nucleosynthesis.}
\end{figure}

\subsection{The Five--Dimensional Picture}
Having applied the four--dimensional effective theory to cosmology, 
we turn now our attention to the five--dimensional interpretation of 
the solutions above. Specifically, we consider the brane positions, 
which are related with $\phi$ and $R$ via eqns. (\ref{posia1}), 
(\ref{posia2}), (\ref{posib1}) and (\ref{posib2}). They are 
shown in fig. 10. It becomes clear 
from eq. (\ref{posib2}), that $R=0$ corresponds to $\sigma =
1/4k\alpha^2$, i.e. the negative tension brane is attracted towards the 
singularity. Negative values of $R$ do not make sense in this 
description, when the transformations (\ref{posia1}), 
(\ref{posia2}), (\ref{posib1}) and (\ref{posib2}) are applied. 
Therefore, cosmological solutions based on the moduli space 
approximation in which $R$ is negative {\it do not 
have a sensible five--dimensional interpretation.} 

\begin{figure}[!ht]\label{positions}
\hspace{5cm}
\psfig{file=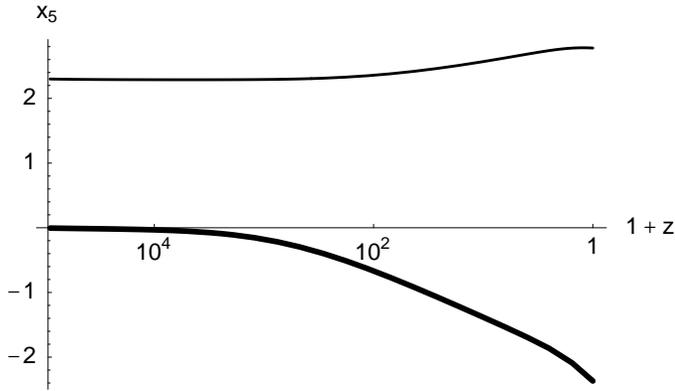,width=9cm}
\vspace{0.75cm}
\caption[h]{Evolution of the brane positions for the cosmology 
as in figure 5.}
\end{figure}

There are two obvious interpretations of this result:

\begin{itemize}
\item At the singularity (i.e. at $R=0$) the negative tension brane 
will be destroyed. There is only one remaining scalar degree of 
freedom $\phi$. The late time evolution is essentially
that of a one brane system. 

\item The negative tension brane is repelled from the singularity. 
 It is tempting to speculate that the repelling can be described by an effective 
potential in the effective four--dimensional description. This option
will be briefly discussed below. 
\end{itemize}

Thus, for a complete understanding of brane cosmology with a bulk scalar 
field and two branes one needs a better understanding of the 
singularity, which  might appear from string theory 
(for a discussion on singularities in string theory see 
e.g. [\ref{Horowitz}] and [\ref{Seiberg}]). 
Nevertheless, we would like to emphasize that the net effect 
remains the same: $R$ is initially attracted towards small values. 

\subsection{Avoiding the Singularity?}
Given that during the matter dominated epoch the negative tension
brane is attracted towards a singularity, we will now investigate the response
of the system with the adoption of the second option mentioned above; i.e. 
we add a potential which drives the branes away from the singularity. 
What are the conditions for such a potential? First, whatever 
type of matter we have on the branes, near the singular point, i.e. 
$Q=0$ ($\phi=-\infty$) and/or $R=0$, the potential term in the 
Klein--Gordon equations must dominate. This can be achieved by an 
inverse--power law potential. A second ingredient is that for 
consistency with the limit $\alpha \rightarrow 0$, 
the derivative of the potential must be proportional to $\alpha^2$ or
some higher power (see eq. (\ref{Qeq})). A simple example 
of a potential which fulfills these requirements is 
\begin{equation}\label{avoid}
V_{\rm add} = V_0 (Q^{\alpha^2} R)^{-\gamma}.
\end{equation}
However, once the negative tension brane is repelled from the
singularity, the field $R$ grows again and is at the present epoch, 
generically, too large to be consistent with observations. Therefore, 
adding a potential which repels the branes from the singularity 
jeopardizes the valuable properties of the attractor mechanism during
the matter dominated epoch. This can be easily understood: the attractor
mechanism is due to the coupling functions $\alpha^{(1)}_R$ and 
$\alpha^{(2)}_R$ appearing in the Klein--Gordon equation (\ref{Rcos}). If at 
some point during the matter dominated epoch these terms become 
negligible compared to the potential term, the field $R$ will be 
driven by the gradient of the potential. For the potential above, 
this will be towards larger values of $R$. Then, at some point the
potential term in the Klein--Gordon equation becomes negligible again compared to the
matter--coupling terms, which drive $R$ towards smaller values 
until the potential term dominates again. During this interplay 
between the terms in the Klein--Gordon equation, $R$ will be driven 
towards larger and larger values because the density of matter drops
and it takes more and more time for the potential term to be
negligible. The expansion of the universe is never dominated by the 
potential energy of the moduli fields for the potential
(\ref{avoid}). 

What happens in the dark energy dominated regime? This depends clearly on the
nature of the dark energy. If the fields $\phi$ and $R$ are responsible 
for dark energy, then we need to modify the potential at large $R$,
as otherwise $R$ will be too large to be consistent with observations 
and, as explained above, the potential (\ref{avoid}), for example,
will not lead to a accelerating universe.
An example which delivers an accelerating universe would be the
potential (\ref{poti}) added to (\ref{avoid}). 
However, this corresponds to the situation where the total potential
has a minimum at $R\ll 1$, which has to be fine--tuned. A cyclic
universe can be modeled by the inclusion of a potential which is
negative in value for some range of the moduli fields [\ref{Turok2}]. 

If instead the dark energy has a different origin and is a field which
is confined on a brane, then there is an important
interplay between the potential term and the coupling term in 
(\ref{Rcos}). We will investigate this possibility in future work.

\section{Conclusions}
In this paper we have investigated the cosmological evolution of brane 
world moduli for a general class of brane world models with two 
boundary branes and a bulk scalar field. The theory we have used 
contains  the
Randall--Sundrum model as a special case. The parameter $\alpha$ 
regulates how strongly the bulk geometry is warped; small values 
of $\alpha$ correspond to highly warped geometries, whereas 
large values of $\alpha$ correspond to only slightly warped 
geometries. 

We have obtained the four--dimensional effective theory from the 
five--dimensional theory using the moduli space approximation. As we 
are interested in late time cosmology, namely from nucleosynthesis on, 
the moduli space approximation should be accurate enough. 
We have discussed observational constraints and found that 
the parameter $\alpha$ must be small, pointing towards a 
warped bulk geometry. Furthermore, the field $R$ must be 
small enough today, meaning that the distances between the 
branes must be large enough. 

As long as matter is not directly coupled to the bulk scalar field, 
and therefore to $\phi$ and $R$, the theory predicts that coupling 
constants such as the fine--structure constant are not space--time 
dependent. Thus, if the observations made in [\ref{webb1}] and 
[\ref{webb2}] hold, more complicated models of brane worlds have to 
be investigated. 

Finally, we have discussed the cosmological evolution of the moduli
fields. We focused in particular on the question of whether $R$ is driven 
towards small values during the cosmological evolution in order 
to be consistent with observations today. 
Our findings indicate that there are cosmological attractors 
for the field $R$, similar to the attractor found in scalar--tensor 
theories [\ref{Damour1}-\ref{Wagoner}]. We have also found that the 
efficiency of the attractor strongly depends on the matter content 
on the negative tension brane. In the five--dimensional picture, 
the attractor solution drives the negative tension brane towards a 
bulk singularity. When $R=0$, the negative tension brane 
hits the singularity. Negative values of $R$ correspond to the 
situation where the singularity is between the branes. As the 
four--dimensional solution suggests an oscillating behaviour of 
$R$ around 0, one has to be careful when interpreting this 
behaviour. Indeed, a proper understanding of the singularity 
is absolutely necessary in order to fully understand the cosmology 
of the two brane system. In the case with the brane repelled 
by the singularity we have argued that in this case the details of the 
evolution of the two brane system can be significantly altered. 

We would like to 
emphasize again that the class of brane world models we have considered in 
this paper predict the coupling functions (\ref{coupling1}) and 
(\ref{coupling2}). These functions control the evolution of the fields 
$\phi$ and $R$ and the coupling to the different matter species.

\vspace{0.5cm}\noindent {\bf Acknowledgements:} We are grateful for 
discussions about the moduli space approximation with T. Wiseman and 
for discussions on multi--scalar tensor theories with 
Gilles Esposito--Farese.
This work was supported by PPARC (C.v.d.B., A.-C.D. and C.S.R.), 
a CNRS--Royal Society exchange grant for collaborative research 
and the European network (RTN), HPRN--CT--200-00148 and 
PRN--CT--2000--00148.

\references
\item \label{Lukas1}
A. Lukas, B.A. Ovrut, K.S. Stelle and D. Waldram, 
Phys.Rev. D{\bf 59}, 086001 (1999)

\item \label{Lukas2} A. Lukas, B.A. Ovrut and D. Waldram, 
Phys.Rev. D{\bf 60}, 086001 (1999)

\item \label{Lukas3} A. Lukas, B.A. Ovrut and D. Waldram, 
Phys.Rev. D{\bf 61}, 023506 (2000)

\item \label{Barreiro1} T. Barreiro and B. de Carlos, 
JHEP 0003, 020 (2000)

\item \label{Barreiro2} T. Barreiro, B. de Carlos and N.J. Nunes, 
Phys.Lett. B{\bf 497}, 136 (2001)

\item \label{Skinner} A. Lukas and D. Skinner, 
JHEP 0109, 020 (2001)

\item \label{Lukas4} M. Brandle, A. Lukas and B.A. Ovrut, 
Phys.Rev. D{\bf 63}, 026003 (2001)

\item \label{RS1} L. Randall and R. Sundrum, 
Phys.Rev.Lett. {\bf 83}, 3370 (1999)

\item \label{RS2} L. Randall and R. Sundrum, 
Phys.Rev.Lett. {\bf 83}, 4690 (1999)

\item \label{Bine1} P. Binetruy, C. Deffayet and D. Langlois, 
Nucl.Phys. B {\bf 565}, 269 (2000)

\item \label{Bine2} P. Binetruy, C. Deffayet, U. Ellwanger and D. Langlois, 
Phys.Lett. B {\bf477}, 285 (2000)

\item \label{Cline1} J.M. Cline, C. Grojean and G. Servant,
Phys.Rev.Lett. {\bf 83}, 4245 (1999)

\item \label{Csaki1} C. Csaki, M. Graesser and J. Terning, 
Phys.Lett.B {\bf 462}, 34 (1999)

\item \label{kribs} C. Csaki, M.L. Graesser and G.D. Kribs, 
Phys. Rev. D{\bf 63}, 065002 (2001)

\item \label{Flana} E.E. Flanagan, S.H.H. Tye and I. Wasserman, 
Phys.Rev. D{\bf 62}, 044039 (2000)

\item \label{Tye} H. Stoica, S.H.H. Tye and I. Wasserman, 
Phys.Lett. B{\bf 482}, 205 (2000)

\item \label{Maartens} R. Maartens, D. Wands, B.A. Bassett and I. Heard,
Phys.Rev. D{\bf 62}, 041301 (2000)

\item \label{Copeland} E.J. Copeland, A.R. Liddle and J.E. Lidsey, 
Phys.Rev. D {\bf 64}, 023509 (2001)

\item \label{Paddy} P.R. Ashcroft, C. van de Bruck and A.-C. Davis, 
astro-ph/0208411

\item \label{Csaki2} C. Csaki, M. Graesser, L. Randall and J. Terning, 
Phys.Rev. D{\bf 62}, 045015 (2000)

\item \label{Cline2} J.M. Cline and J. Vinet, JHEP 0202, 042 (2002)

\item \label{Wise1} W.D. Goldberger and M. B. Wise,
Phys.Rev. D {\bf 60}, 107505 (1999)

\item \label{Wise2} W.D. Goldberger and M. B. Wise,
Phys.Rev.Lett. {\bf 83}, 4922 (1999)

\item \label{Garriga1} J. Garriga and T. Tanaka, 
Phys.Rev.Lett. {\bf 84}, 2778 (2000)

\item \label{Chiba} T. Chiba, Phys.Rev.D {\bf 62}, 021502 (2000)

\item \label{Tanaka} T. Tanaka, X. Montes, 
Nucl.Phys.B {\bf 582}, 259 (2000)

\item \label{Bine3} P. Binetruy, C. Deffayet and D. Langlois, 
Nucl. Phys. B{\bf 615}, 219 (2001)

\item \label{Brax1} Ph. Brax, C. van de Bruck, A.-C. Davis and C.S. Rhodes, 
Phys.Rev. D {\bf 65}, 121501 (2002)

\item \label{wiseman} T. Wiseman, Class.Quant.Grav. {\bf 19}, 
3083 (2002)

\item \label{Justin} J. Khoury and R.--J. Zhang, 
Phys.Rev.Lett. {\bf 89}, 061302 (2002)

\item \label{Kanno1} S. Kanno and J. Soda, Phys.Rev.D {\bf 66}, 
043526 (2002)

\item \label{Kanno2} S. Kanno and J. Soda, hep-th/0207029 

\item \label{Choi} K. Choi, H.B. Kim and H.D. Kim,
Mod.Phys.Lett. A {\bf 14}, 125 (1999)

\item \label{Carlos} T. Barreiro, B. de Carlos and E.J. Copeland, 
Phys.Rev. D{\bf 58}, 083513 (1998)

\item \label{Stein} P.J. Steinhardt, Phys.Lett. B {\bf 462}, 41 (1999)

\item \label{Steinhardt} G. Huey, P.J. Steinhardt, B.A. Ovrut and
D. Waldram, Phys.Lett. B{\bf 476}, 379 (2000)

\item \label{Damour1} T. Damour and K. Nordtvedt, 
Phys.Rev.Lett. {\bf 70}, 2217 (1993)

\item \label{Damour2} T. Damour and K. Nordtvedt, 
Phys.Rev.D {\bf 48}, 3436 (1993)

\item \label{Wagoner} D.I. Santiago, D. Kalligas and R. Wagoner, 
Phys.Rev.D {\bf 58}, 124005 (1998)

\item \label{webb1} J.K Webb, V.V. Flambaum, C.W. Churchill, 
M.J. Drinkwater and J.D. Barrow, Phys.Rev.Lett. {\bf 82}, 884 (1999)

\item \label{webb2} J.K. Webb, M.T. Murphy, V.V. Flambaum, V.A. Dzuba, 
J.D. Barrow, C.W. Churchill, J.X. Prochaska and A.M. Wolfe, 
Phys.Rev.Lett. {\bf 87}, 091301 (2001)

\item \label{Will} C. Will, {\it Theory and Experiment in Gravitational 
Physics}, Cambridge University Press (1993)

\item \label{Garriga2} J. Garriga, O. Pujolas and T. Tanaka, 
hep-th/0111277

\item \label{Brax2} Ph. Brax and A.-C. Davis, 
Phys.Lett.B {\bf 497}, 289 (2001)

\item\label{us} Ph. Brax, C. van de Bruck, A. -C Davis and C. -S. Rhodes,
Phys. Lett. B {\bf 531}, 135 (2002) 

\item\label{cynolter} G. Cynolter, hep-th/0209152

\item \label{Turok1} J. Khoury, B.A. Ovrut, P. Steinhardt and 
N. Turok, Phys.Rev.D {\bf 64}, 123522 (2001)

\item \label{Turok2} P. Steinhardt and N. Turok, 
Phys.Rev.D {\bf 65}, 126003 (2002)

\item \label{Wald} R. Wald, {\it General Relativity}, University 
of Chicago Press (1984)

\item \label{Gibbons} T. Damour, G.W. Gibbons and C. Gundlach, 
Phys.Rev.Lett. {\bf 64}, 123 (1990)

\item \label{Gundlach} T. Damour  and C. Gundlach, 
Phys.Rev. D {\bf 43}, 3873 (1991)

\item \label{Brax4} Ph. Brax and J. Martin, 
Phys.Rev. D {\bf 61}, 103502 (2000)

\item \label{gilles} T. Damour and G. Esposito-Farese, 
Class.Quant.Grav. {\bf 9}, 2093 (1992)

\item\label{Uzan} J.--P. Uzan, hep-ph/0205340

\item\label{Chiba2} T. Chiba, gr-qc/0110118

\item \label{Wetterich} C. Wetterich, 
Nucl.Phys. B{\bf 302}, 668 (1988)

\item \label{Peebles} B. Ratra and P.J.E. Peebles, 
Phys.Rev. D{\bf 37}, 3406 (1988)

\item \label{Pedro} P.G. Ferreira and M. Joyce, 
Phys.Rev. D{\bf 58}, 023503 (1998)

\item \label{Binetruy5} P. Binetruy, 
Phys.Rev.D {\bf60}, 063502 (1999)

\item \label{Martin} Ph. Brax and J. Martin, 
Phys.Lett. B{\bf 468}, 40 (1999)

\item \label{Copeland2} E.J. Copeland, N.J. Nunes and 
F. Rosati, Phys.Rev. D{\bf 62}, 123503 (2000)

\item \label{Davis} Ph. Brax and A.C. Davis, 
JHEP 0105, 007 (2001)

\item \label{Bruck} Ph. Brax, C. van de Bruck and A.C. Davis, 
JHEP 0110, 026 (2001)

\item \label{Linde} R. Kallosh, A. Linde, S. Prokushkin and 
M. Shmakova, hep-th/0208156 

\item \label{Horowitz} G. T. Horowitz and J. Polchinski, 
hep-th/0206228 

\item \label{Seiberg} H. Liu, G. Moore and N. Seiberg, 
hep-th/0206182

\end{document}